\documentclass[]{jfm}

\usepackage{graphicx}
\usepackage{newtxtext}
\usepackage{newtxmath}
\usepackage{natbib}
\usepackage{hyperref}
\usepackage{subfig}
\usepackage{xcolor}
\hypersetup{
    colorlinks = true,
    urlcolor   = blue,
    citecolor  = black,
}

\usepackage{siunitx}

\newcommand{\RomanNumeralCaps}[1]

\usepackage[perpage]{footmisc}

\usepackage{perpage}

\title{Breakup of viscous liquid bridges on solid surfaces}

\author{{Salar Farokhi\aff{1}},
Peyman Rostami \aff{2},
G\"unter K. Auernhammer \aff{2},
Steffen Hardt \aff{1},
 \corresp{\email{hardt@nmf.tu-darmstadt.de}}}

\affiliation{\aff{1}  Fachgebiet Nano- und Mikrofluidik, TU Darmstadt, Darmstadt, Germany
\aff{2}Leibniz-Institut f\"ur Polymerforschung Dresden e.V., Dresden, Germany}

\begin{document}
\maketitle

\begin{abstract}
The breakup dynamics of viscous liquid bridges on solid surfaces is studied experimentally. 
It is found that the dynamics bears similarities to the breakup of free liquid bridges in the viscous regime. 
Nevertheless, the dynamics is significantly influenced by the wettability of the solid substrate. Therefore, it is essential to take into account the interaction between the solid and the liquid, especially at the three-phase contact line.
It is shown that when the breakup velocity is low and the solid surface is hydrophobic, the dominant channel of energy dissipation is likely due to thermally activated jumping of molecules, as described by the Molecular Kinetic Theory.
Nevertheless, the viscous dissipation in the bulk due to axial flow along the bridge can be of importance for long bridges. In view of this, a scaling relation for the time dependence of the minimum width of the liquid bridge is derived. 
For high viscosities, the scaling relation captures the time evolution of the minimum width very well. Furthermore, it is found that external geometrical constraints alter the dynamic behavior of low and high viscosity liquid bridges in a different fashion. 
This discrepancy is explained by considering the dominant forces in each regime. Lastly, the morphology of the satellite droplets deposited on the surface is qualitatively compared to that of free liquid bridges.

\end{abstract}


 \section{Introduction}
\label{sec:introduction}

The standard scenario of capillary bridge breakup is a liquid bridge spanning the space between two parallel surfaces, here referred to as “free liquid bridge”. 
Over the past decades, free liquid bridges have been in the focus of intense research activities. Three main regimes of liquid bridge breakup have been identified, a viscous regime, an inertial regime and a viscous-inertial regime (\cite{li2016capillary}).
In the viscous regime, the dynamics is governed by a balance between capillary and viscous forces, while in the inertial regime, the balance is between capillary and inertial forces. In the viscous-inertial regime, viscous and inertial forces are of equal magnitude. 
For these different breakup regimes, scaling relationships and similarity solutions have been derived, see, e.g., (\cite{brenner1996pinching, eggers2012stability, eggers2008physics, eggers1993universal, papageorgiou1995analytical, papageorgiou1995breakup} ). 
In the past few years, these studies have been extended to more complex scenarios, e.g., to viscoelastic liquids (\cite{chen2021viscoelastic}), or to bridges through which an electric current flows (\cite{pan2021armstrong}).
Much less attention has been spent on the breakup of liquid bridges that extend along solid surfaces, in the following referred to as “wetting capillary bridges”. 
Such structures can be found on surfaces with regions of different wettability. Drop deposition on wettability patterns is relevant for multilayer inkjet printing, where different layers of ink are printed above each other. This technology is used for the fabrication of fuel cells (\cite{towne2007fabrication}), solar panels (\cite{el2021overview}) or electric circuits (\cite{kang2012all, kawamoto2007electronic, wang2016based}), among others. 
When a drop deposited on a wettability pattern evaporates, liquid bridges can be found on a hydrophobic region between two hydrophilic regions (\cite{hartmann2019stability}). When the liquid volume gets reduced by evaporation, the liquid bridge becomes thinner and thinner until a point of instability is reached, after which rapid bridge breakup is observed (\cite{hartmann2019stability}). 
Numerical simulations of the breakup dynamics of water bridges have shown that the process is mainly controlled by a balance between inertial and capillary forces (\cite{hartmann2021breakup}).

\begin{figure}%
    \centering
    \subfloat[\centering ]{{\includegraphics[width=6cm]{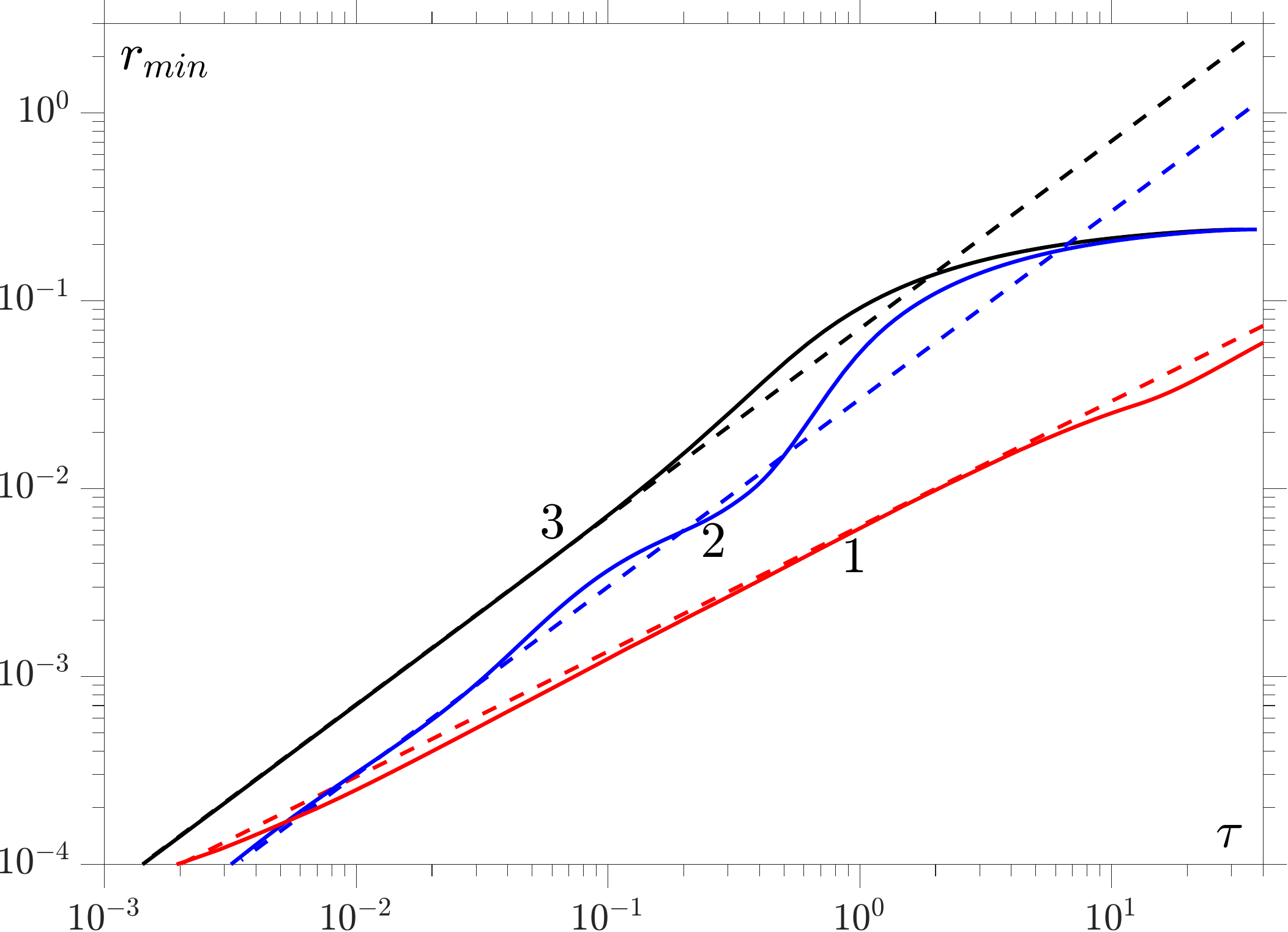} }}%
    \qquad
    \subfloat[\centering ]{{\includegraphics[width=6cm]{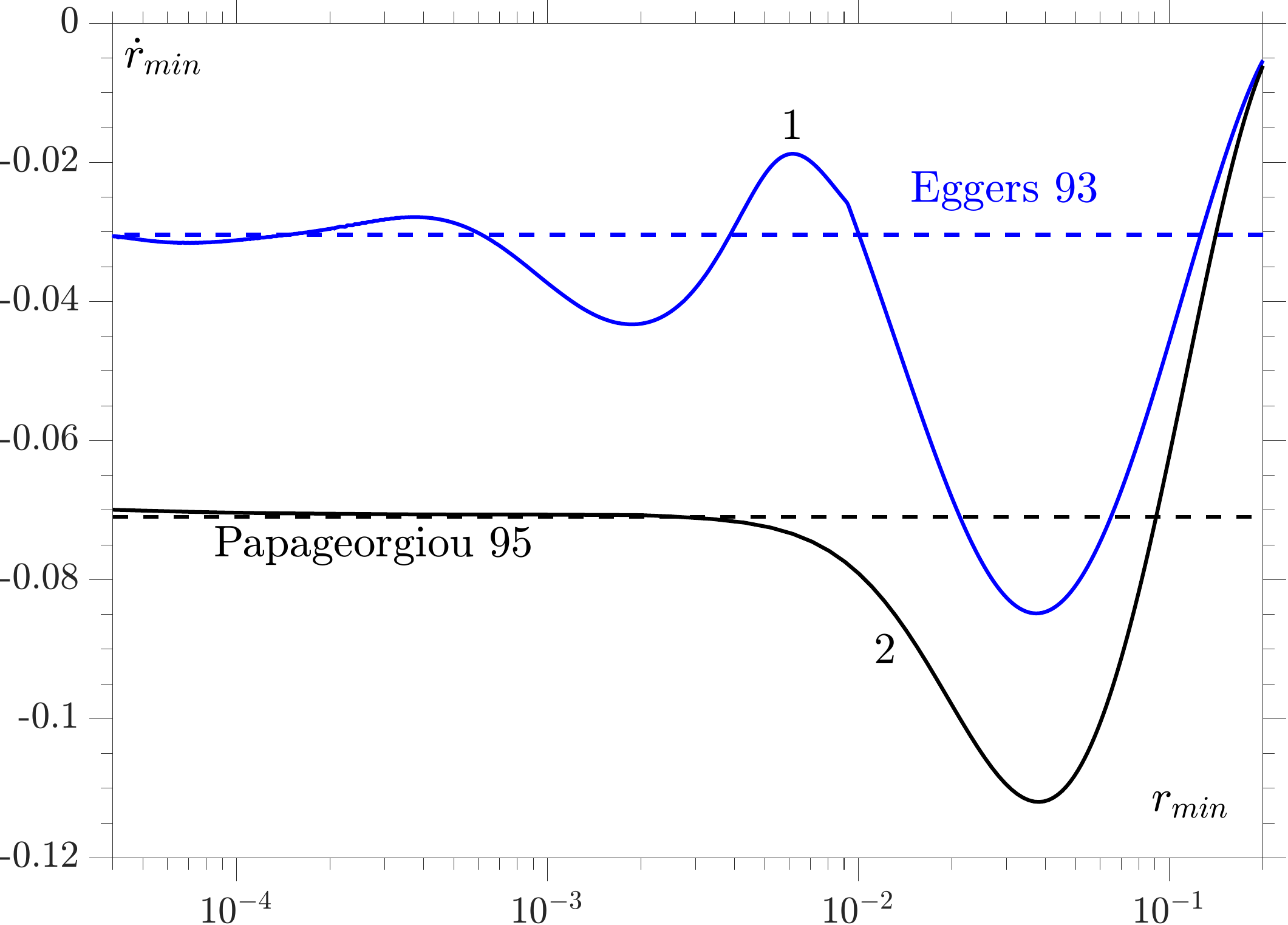} }}%
    \caption{Characteristic time evolution of free liquid bridges before breakup in different regimes. (a) Minimum bridge radius against time in the inertial (1), viscous-inertial (2) and viscous (3) regime. The dashed lines indicate similarity solutions. (b) Breakup speed against minimum bridge radius for the viscous-inertial (1) and the viscous (2) regime. Again, the dashed lines indicate similarity solutions. Reproduced from (2016 Sprittles).  }%
    \label{fig1}%
\end{figure}

The study on the breakup of wetting capillary bridges presented in this paper relies on the same schemes for analyzing data as employed for free liquid bridges. Therefore, subsequently we briefly introduce these schemes, where we largely follow the analysis of the dynamic regimes presented in (\cite{li2016capillary}). When measuring the minimum radius of a free capillary bridge before breakup, usually curves similar to those shown in Figure ~\ref{fig1} are obtained.
The three different curves represent the inertial, viscous-inertial and viscous regime. At least during a part of the time evolution, the curves align with predictions from similarity solutions, which are essentially different power laws for the different dynamic regimes (\cite{li2016capillary}). 
It needs to be emphasized that it cannot be expected that the breakup dynamics remains in the same regime during the entire time evolution. Rather than that, transitions between different regimes will occur (\cite{li2016capillary}). 
For example, when the viscous regime prevails at times relatively long before the point of breakup, a transition to the viscous-inertial regime will occur. However, the transition point may be so close to the point of breakup that the regime transition is hardly measurable or detectable. Apart from the radius of the capillary bridge, the time derivative of the minimum radius, termed “breakup velocity”, is a key quantity characterizing the breakup process. Similarity solutions predict a constant breakup velocity for the viscous and viscous-inertial regime (\cite{li2016capillary}) when plotted as a function of the minimum bridge radius, which is exemplified in Figure ~\ref{fig1}.b.

In this manuscript, we will use similar methods to characterize the breakup dynamics of viscous wetting capillary bridges. 
We employ high-speed imaging to study the dynamics of liquids of different viscosities on different surfaces and show that for most liquids, the dynamics follows a viscous scaling law. 
A marked difference to the breakup dynamics of free liquid bridges lies in the fact that for wetting liquid bridges, a very significant contribution to viscous dissipation is due to the three-phase contact line.

\section {Experimental method}

\subsection{Microfluidic setup}

\begin{figure}
\centering
\includegraphics[width=0.75\linewidth]{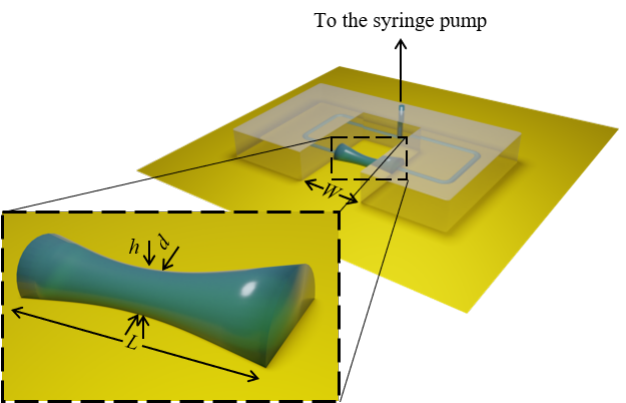}
\caption{Schematics of the microfluidic setup, and magnified view of the capillary bridge}
\label{fig2}
\end{figure}

Figure ~\ref{fig2} shows the experimental setup used to create and manipulate the capillary bridge on the substrate. 
The setup consists of a polydimethylsiloxane (PDMS) block attached to a substrate (without plasma treatment of the surface or the PDMS block) and connected to a syringe pump (KDS Legato 200). 
The PDMS block is fabricated using a standard soft lithography method. 
After thoroughly mixing the PDMS (Sylgard 184, Dow, Germany) with the corresponding crosslinker in a 10:1 ratio, the mixture is degassed in a desiccator for 1 h to remove the gas bubbles formed during mixing. Subsequently, the mixture is poured onto the 3D-printed master structure (Prusa SL1S 3D printer) and cured in an oven at \SI{40}{\celsius} for 24 h.
The resulting structure has two channels with a semi-circular cross-section and a diameter of \SI{500}{\micro \meter}. 
The gap width (W) varies between 0.75 and \SI{3}{\milli \meter}.

\subsection{Substrates and liquids} \label{section 2.2}

The substrates are prepared as follows.
Silicon wafers (CZ-Si wafer, 4-inch diameter, thickness 500 $\pm$ \SI{50}{\micro \meter}, Microchemicals GmbH, Germany) are first coated with a \SI{15}{n \meter} layer of chromium, and subsequently, a \SI{50}{n \meter} layer of gold using an E-beam method (Balzers BAK 600).
Before the coating process, the chamber is preheated to \SI{75}{\celsius} and evacuated to a pressure of $6.7\times 10^{-6}$ \si{\milli \bar}. 
During the coating, the chamber temperature and pressure are around \SI{85}{\celsius} and $2\times 10^{-5}$ \si{\milli \bar}. In order to reach coating rates of 0.4 and  \SI{0.2}{n\meter \per \second} for chromium and gold, respectively, \SI{5}{\kilo \volt} is applied to evaporate the mentioned materials. 
The chromium/gold-coated silicon wafers are used as the more hydrophilic substrates. 
We will refer to this type of surface as “Au”. The advancing and receding contact angle of water on this surface is \SI{96}{\degree} and \SI{67}{\degree}, respectively, determined with a Krüss DSA 100 Drop Shape Analyzer. 
Gold substrates are known to have high surface energy and consequently low contact angles (\cite{white1964wetting, bewig1965wetting}). 
The main reason for this difference is the contamination of the substrate with impurities. 
To support our hypothesis, we placed the substrates on a hotplate at \SI{250}{\celsius} for 6 minutes, which reduced the contact angle to \SI{35}{\degree}. 
For the rest of this work, we use the chromium/gold coated silicon wafer without any further heat treatment. To achieve more hydrophobicity, a thin PDMS pseudo brush layer is applied to the surfaces. 
Similar coating processes are detailed elsewhere (\cite{eifert2014simple}), but a brief summary is provided here. 
First, the surfaces are cleaned in an $O_{2}$ plasma chamber (Diener Femto). 
Next, \SI{0.5}{\milli \liter} of a 5 cSt silicone oil (CAS: 63148-62-9, Sigma-Aldrich, Germany) is dispensed on the surface with a pipette, and the surface is pre-heated to \SI{100}{\celsius}. 
After the silicone oil spreads over the entire surface, the surface is placed on a hot-plate at \SI{220}{\celsius} for three minutes. 
Finally, the substrates are rinsed with isopropanol and water for 10 seconds each. 
We refer to this type of surface as “PDMS$@$Au”. The advancing and receding contact angle of water on this surface is \SI{109}{\degree} and \SI{95}{\degree}, respectively, i.e. it is the more hydrophobic one of the two. 
Since the focus of this study is on viscosity variations, mixtures of water (purified using a Milli-Q device; specific resistance \SI{18.2}{\mega\ohm.\centi\meter} at \SI{25}{\celsius}) and glycerol ($> $99.5\%, Sigma-Aldrich, Germany) with different glycerol contents are chosen as the working liquids. 
The viscosities of the mixtures, measured with a rheometer (Brookfield DV-III Ultra), and the literature values of the surface tension (\cite{takamura2012physical}) are presented in Table~\ref{Liquids properties}, where the percentage represents the mass of glycerol relative to the mass of the mixture. 
Throughout the remainder of this article, liquids will be referred to by their glycerol percentage.

\begin{table}
\centering
\begin{tabular}{ccc}
Liquid  & Viscosity (\SI{}{\milli\pascal\second}) & Surface tension (\SI{}{\milli\newton \per \meter}))     \\
\hline
0\% & 1 & 71.68  \\
30\% &2.5     & 70.38 \\
50\% &5.9 &  69.02 \\
60\% & 10.9  & 68.11  \\
70\% & 23.7 & 66.97  \\
\hline
\end{tabular}
\caption{\label{Liquids properties} Viscosities of the water-glycerol mixtures used in this study as a function of glycerol content.} \end{table}

\subsection{Experimental procedure}

To create the capillary bridge, liquid is pumped through the channels by a syringe pump.
The two streams meet on the surface and shape the capillary bridge, after which the pumping is stopped. 
Next, to bring the capillary bridge to the unstable configuration, liquid is withdrawn by the syringe pump. 
In order to prevent any bias of the breakup dynamics due to the withdrawal, the withdrawal rate is kept as low as \SI{2}{\micro \liter / \min}.
It is known that cured PDMS can contain uncrosslinked chains that diffuse to its surface and potentially alter the wetting behavior (\cite{jensen2015wetting}). 
However, these polymer chains have a very low solubility in aqueous solutions and the results obtained with the current setup for water closely resemble those from a previous study(\cite{hartmann2021breakup}), in which a completely different setup was used.
Therefore, the effect of such polymer chains on the dynamics is marginal. In all experiments, the gold layer is grounded to exclude effects due to slide electrification (\cite{ratschow2024charges,li2023surface,li2022spontaneous}) .
A coaxial imaging system with a high-speed camera (Photron FASTCAM SA-1.1) in conjunction with 12X macro objective (Navitar) is used to observe the system from the top view. 
Finally, an in-house MATLAB script with a sub-pixel edge detection method based on a work by (\cite{trujillo2013accurate}) with a maximum position error of around 0.65 pixels was developed to extract the minimum width of the capillary bridge as well as its breakup velocity from the recorded images. 
Subject to the required field of view and thus magnification, the pixel resolution varies between 3.66 and \SI{1.57}{\micro \meter}. 
Depending on the speed of the process, the frame rate was varied between 20,000 and 54,000 frames per second (fps). 
Due to the limited temporal resolution, it is not possible to determine the breakup time arbitrarily accurately. 
Therefore, similar to (\cite{hartmann2021breakup}), the breakup is assumed to occur $\Delta t/2$ seconds ($\Delta t$ being the time interval between two consecutive frames) before the first recorded image where the liquid bridge appears to be broken up.
The time before breakup is defined as $\tau = t_{b}-t $, where $t_{b}$ is the breakup time. 
For instance, for the first frame where the liquid is broken up, we have $\tau = -\Delta t/2$, whereas in the final recorded image where the liquid bridge remains intact we have $\tau = \Delta t/2$.

\section {Results }

 \subsection{Influence of the solid substrate}\label{section3.1}
 
\begin{figure}
\centering
\includegraphics[width=0.75\linewidth]{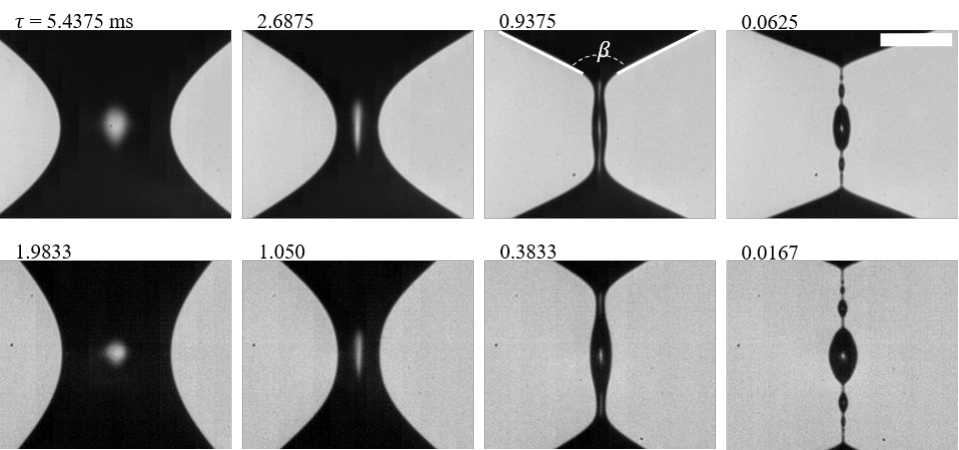}
\caption{Snapshots of the time-evolution of the capillary bridge for the 60\% liquid on the Au (top row) and the PDMS$@$Au (bottom row) surface. $\tau$ indicates the time before breakup for each image in ms. The scale bar represents \SI{100}{\micro \meter}.}
\label{fig3}
\end{figure}

This section is dedicated to the effect of substrate wettability on the dynamics of the capillary bridge. 
To this end, the two surfaces referred to as Au and PDMS$@$Au are used. Snapshots of the evolution of the capillary bridge for the 60\% liquid on the aforementioned surfaces are shown in Figure ~\ref{fig3}. 
The dark regions correspond to the capillary bridge (with light reflection in the middle), while the brighter regions correspond to the solid surface. 
The frames are selected in a way that the minimum width of the bridges ($d$) for the two cases in each column is equal. 
This implies that the time before breakup $\tau$ (see section ~\ref{section 2.2}) is different for two corresponding frames in the top and bottom rows. 
For example, the first frame for the PDMS$@$Au case corresponds to approximately \SI{1.98}{\milli \second} before breakup, whereas that of the Au case it corresponds to around \SI{5.44}{\milli \second}.
It should be noted that the position of the minimum width during breakup shifts. 
This point will be further discussed in section ~\ref{section3.4}.

\begin{figure}%
    \centering
    \subfloat[\centering ]{{\includegraphics[width=6cm]{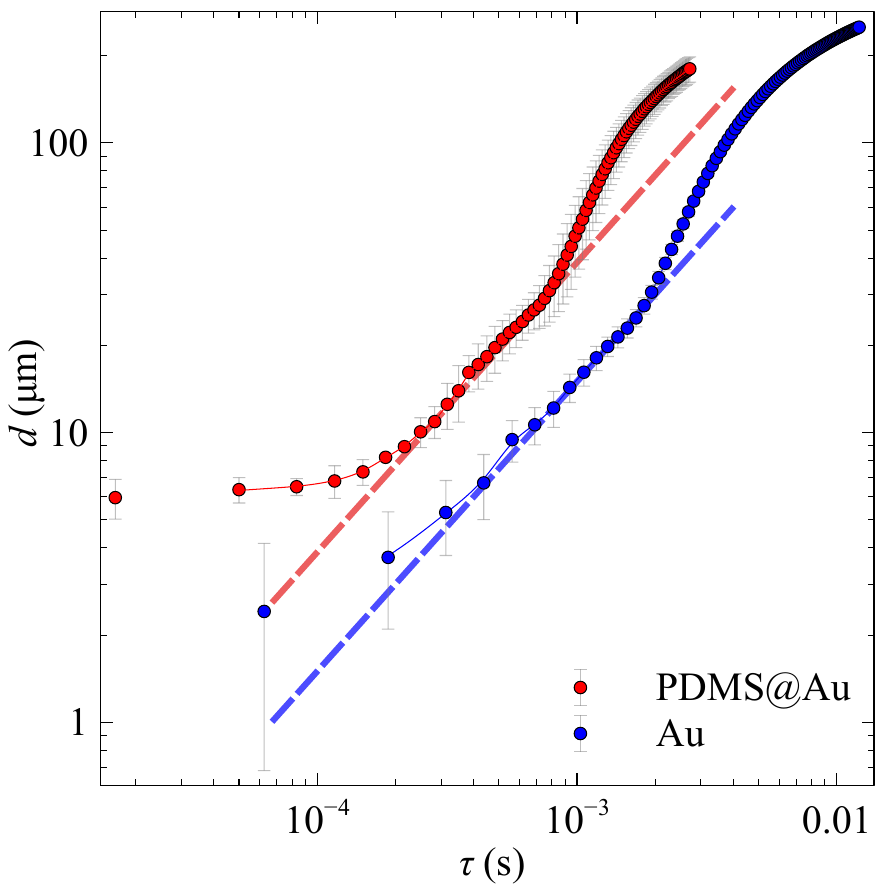} }}%
    \qquad
    \subfloat[\centering ]{{\includegraphics[width=6cm]{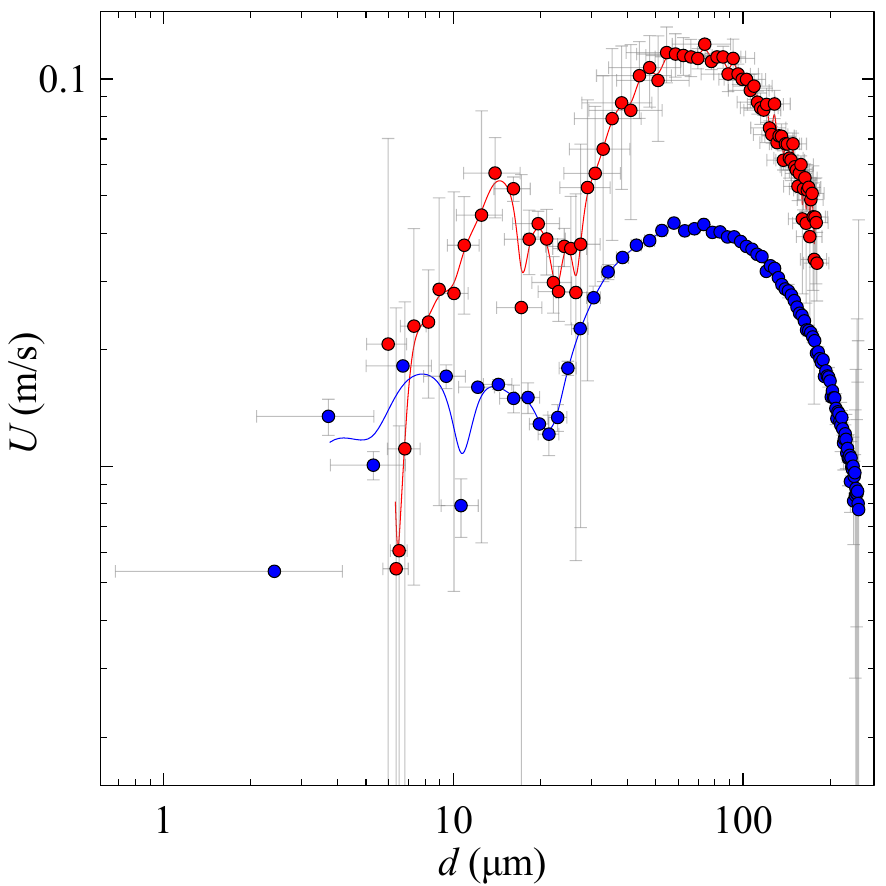} }}%
    \caption{Influence of the solid substrate on the dynamics of the wetting capillary bridge for the 60\% liquid: (a) Minimum width vs. time before breakup, (b) breakup velocity vs. minimum width. The symbols represent measurements, and the solid lines are B-spline fits. The dashed lines represent power laws with a power of one. The error bars indicate the standard deviation. The legend of subfigure (a) applies to all the subfigures. }%
    \label{fig4}%
\end{figure}

Clearly, the two cases exhibit differences in shape, such as the cone opening angle ($\beta$) and the axial elongation of the capillary bridge.
These differences are also visible in the $d$ vs. $\tau$ and $U$ ($=\frac{\mathrm{d} d}{\mathrm{d} \tau }$) vs. $d$ data, see Figure ~\ref{fig4}.a and Figure ~\ref{fig4}.b.
The symbols represent measurements, and the solid lines depict B-spline fits to the data points. 
In this and in some of the following figures, the B-spline fits merely serve as guides to the eye. 
As shown in Figure ~\ref{fig4}, qualitatively, in each case the dynamic behavior is very similar to the dynamic behavior of highly viscous free capillary bridges (\cite{li2016capillary, papageorgiou1995breakup}) as indicated by the convergence of the curves to the power laws with a power of one (the dashed lines). 
However, there is a substantial dependence of the dynamics on surface properties. 
For instance, the breakup velocity on the PDMS$@$Au (more hydrophobic) surface is much higher (approximately three times higher on average) than on the Au surface. 
It should be mentioned that a similar behavior –namely, faster breakup dynamics on PDMS$@$Au surfaces– is also observed with other liquids (data not shown). 
Therefore, it can be inferred that the more hydrophobic surface imposes less resistance on the motion of the capillary bridge than the Au surface, leading to higher velocity and greater elongation in the axial direction for the former case. 
The dependence of the dynamics on the wetting properties implies that the solid-liquid interaction, especially at the three-phase contact line, must be taken into account to understand the dynamics of capillary breakup of viscous liquids on solid surfaces.

\subsection{Influence of viscosity}

\begin{figure}
\centering
\includegraphics[width=0.75\linewidth]{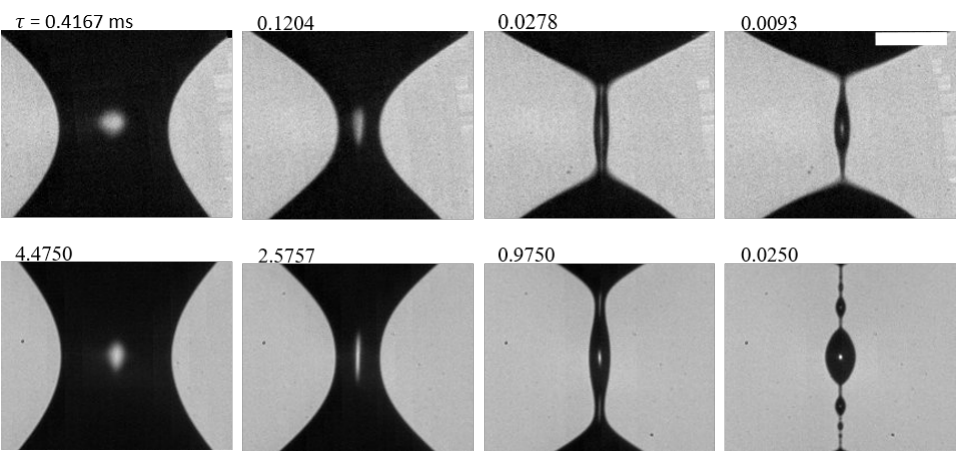}
\caption{Snapshots of the time-evolution of the capillary bridge for the 0\% liquid (top row) and the 70\% liquid (bottom row) on the PDMS$@$Au surface. $\tau$ indicates the time before breakup for each image in ms.  The scale bar represents \SI{100}{\micro \meter}.}
\label{fig5}
\end{figure}

To clarify the influence of viscosity on the dynamic behavior of the capillary bridge on the solid surfaces, five different water-glycerol mixtures were used in the experiments. 
Figure ~\ref{fig5} depicts the time evolution of the capillary bridge for the 0\% and 70\% liquids. 
The frames are selected in a way that the minimum width of the bridges ($d$) for the two cases in each column is comparable. 
All experiments were conducted on the PDMS$@$Au surface. 
As can be seen in the snapshots, the more viscous liquid bridge has a larger axial elongation, as well as a smaller cone opening angle.

\begin{figure}%
    \centering
    \subfloat[\centering ]{{\includegraphics[width=6cm]{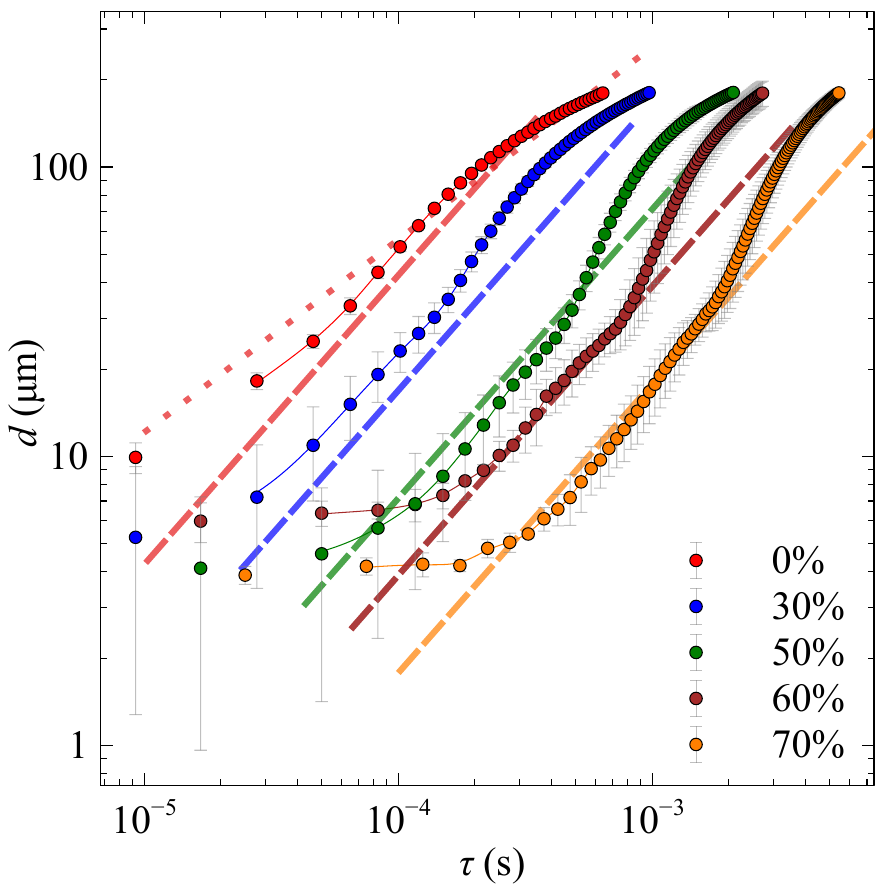} }}%
    \qquad
    \subfloat[\centering ]{{\includegraphics[width=6cm]{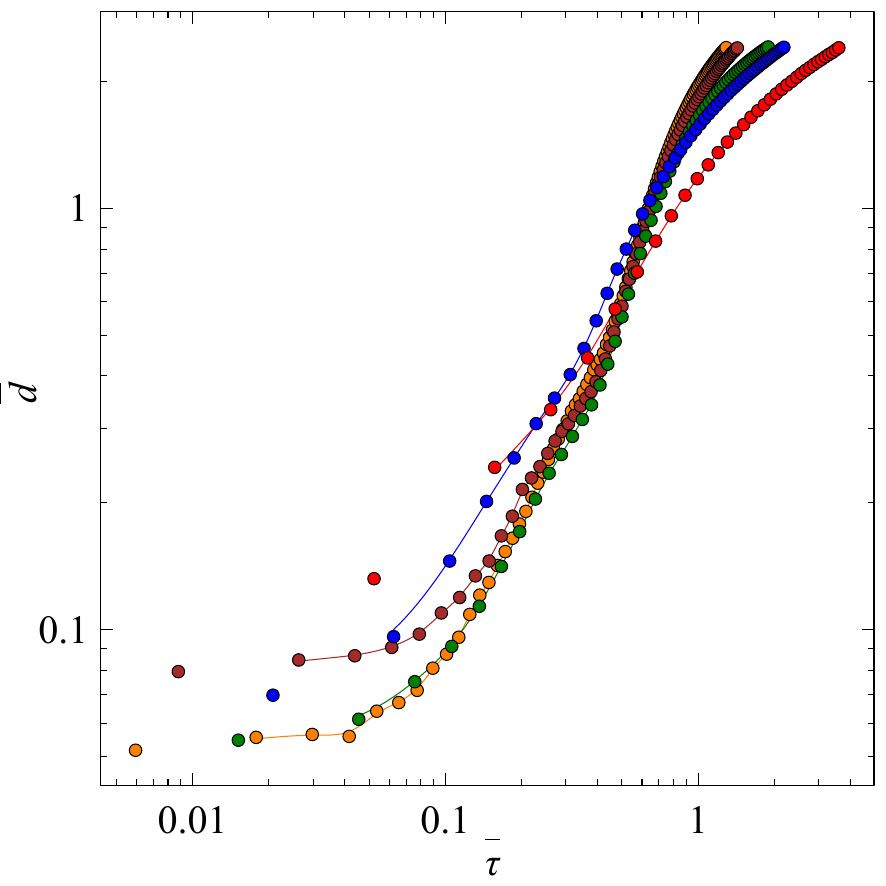} }}%
    \qquad
    \subfloat[\centering ]{{\includegraphics[width=6cm]{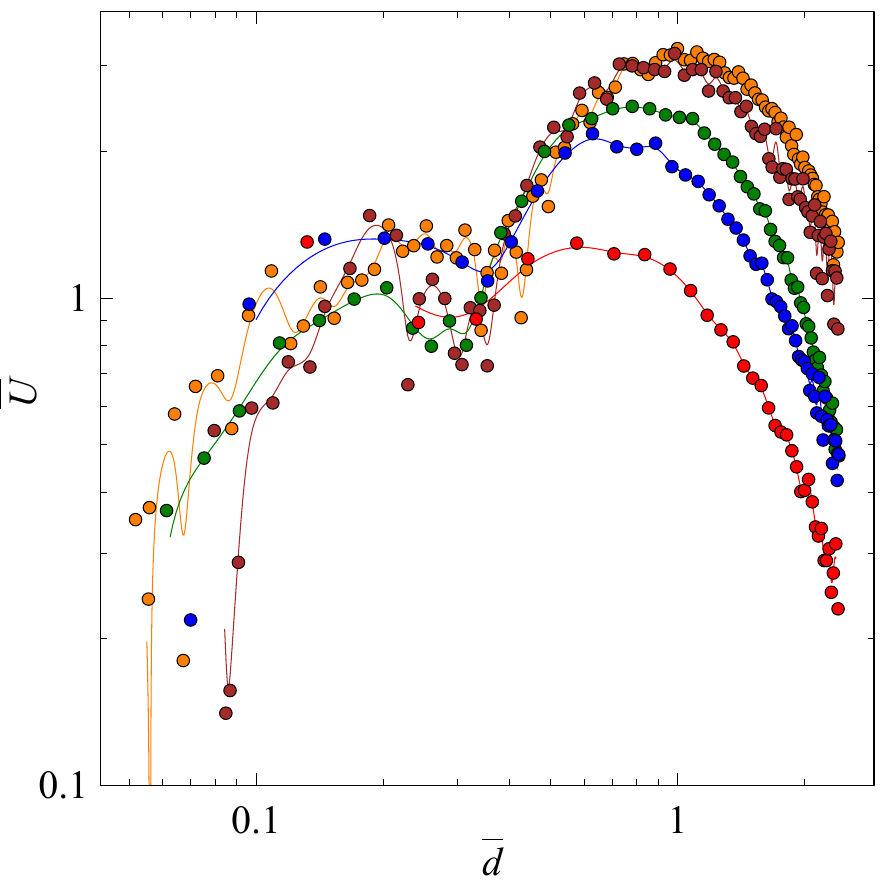} }}
    \caption{Influence of viscosity on the dynamics of the capillary bridge on the PDMS$@$Au surface. (a) Minimum width vs. time before breakup. The dashed lines represent power laws with a power of one. The error bars represent the standard deviation. (b) Scaled minimum width vs. scaled time before breakup. (c) Scaled breakup velocity vs. scaled minimum width. The legend of subfigure (a) applies to all the subfigures.}%
    \label{fig6}%
\end{figure}

Figure ~\ref{fig6}.a shows the $d$ vs. $\tau$ plots for various viscosities. 
The symbols represent measurements and the solid lines are B-spline fits to the data points. 
The dashed lines represent a power law with a power of one, whose relevance will be further discussed in this section (for more details, see ~\ref{section3.2.2}). 
It can be seen that the dynamics of the 50\%-70\% liquids look very similar, the curves are merely shifted along the x-axis. 
Keeping in mind that Figure ~\ref{fig6}.a is a log-log plot, the horizontal shift of the curves represents a change in the velocity ($U=\frac{\mathrm{d} d}{\mathrm{d} \tau }$).
In other words, by increasing the viscosity, the breakup process slows down. 
It should be mentioned that the breakup dynamics for the case of low viscosity liquids (e.g. water) has been investigated by (\cite{hartmann2021breakup}). 
In that study, it was concluded that the dynamics follows an inertial scaling $d\sim \tau ^{2/3}$ in an average sense. 
This is also confirmed in the present study, indicated by the fact that the red dotted line has a slope of 2/3.
Considering the similar qualitative behavior of highly viscous free and wetting capillary bridges, one might be tempted to assume the viscous scaling for the free capillary bridge to be valid for the capillary bridge on a solid substrate. 
However, as already noted in Section ~\ref{section3.1}, the scaling for the latter must take into account the wetting properties of the surface.

\subsubsection{Dominating sources of dissipation }

The motion of a three-phase contact line is an irreversible process, resulting in energy dissipation (\cite{blake2006physics, de1985wetting}). 
There are two common ways of describing the energy dissipation at moving contact lines: the so-called hydrodynamic theory and the molecular kinetic theory (MKT). 
The hydrodynamic theory considers the viscous dissipation in the wedge near the contact line based on the Navier-Stokes equation (\cite{bonn2009wetting, voinov1976hydrodynamics}). 
Since this dissipation is a function of the wedge angle (dynamic contact angle), surface wettability enters the equations. 
It has been shown that (\cite{bonn2009wetting}):

\begin{equation}\label{Eq3.1}
D_{hyd}\approx \eta U^{2}\frac{\ln (\frac{R_{0}}{R_{i}})}{\theta _{D}}L,
\end{equation}

where $D_{hyd}$ is the dissipated power according to the hydrodynamic theory, $\eta$ is the liquid viscosity, $U$ is the contact line velocity, $R_{i}$ is a microscopic cut-off length, $R_{0}$ is a macroscopic length scale (drop radius, for example), $\theta _{D}$ is the dynamic contact angle, and $L$ is the length of the contact line.
By contrast, the MKT neglects the hydrodynamic viscous dissipation altogether and introduces a molecular mechanism for the dissipation at the contact line (\cite{blake2002influence}). 
According to this theory, molecular motion in the vicinity of the contact line is the dominant channel of dissipation. 
It has been shown that the contact line velocity is related to the contact angle through the following relation:

\begin{equation}\label{Eq3.2}
U=2\kappa ^{0}\lambda \sinh \left ( \gamma \lambda ^{2} \frac{\cos (\theta _{S})-\cos (\theta _{D})}{k_{B}T}\right ),
\end{equation}

where $U$ is the contact line velocity, $\kappa ^{0}$ is the frequency at which molecular jumps occur at equilibrium, $\lambda$ is the jump length, $\gamma$ is the surface tension, $\theta _{S}$ and $\theta _{D}$ are the static and dynamic values of the contact angle, $k_{B}$ is the Boltzmann constant, and $T$ is the absolute temperature. 
If the argument of the sinh function is not too large, that is the contact angle does not significantly deviate from its static value, this equation can be linearized to the following:

\begin{equation}\label{Eq3.3}
U=\frac{\gamma }{\zeta } (\cos (\theta _{S})-\cos (\theta _{D})),
\end{equation}

where $\zeta = k_{B}T/\kappa ^{0}\lambda ^{3}$.
It should be noted that for a given surface at constant temperature, this factor is constant and is denoted as “contact line friction”. 
It has been shown that contact line friction can be related to the static contact angle as well as viscosity through the following equation (\cite{blake2002influence, blake2006physics}):

\begin{equation}\label{Eq3.4}
\zeta = \eta (\frac{v_{L}}{\lambda ^{3}})\exp (\frac{\lambda ^{2} Wa}{k_{B}T}),
\end{equation}

where $v_{L}$ is the volume of unit flow, which for many simple liquids is equal to the molecular volume, and $Wa=1+\cos (\theta _{S})$ is the work of adhesion (\cite{blake2002influence}). 
Intuitively, increasing the liquid viscosity or reducing the hydrophobicity of the surface would increase the dissipation at the contact line.

Re-arranging the terms in Equation (3.3), it can be inferred that the surface force $\gamma(\cos (\theta _{S}-\cos (\theta _{D} ))$ on the contact line that is obtained by considering the force balance on a control volume is equal to the friction force on the contact line $\zeta U$; hence the term “contact line friction”. 
Therefore, the dissipated power of the contact line  through this mechanism can be written as:

\begin{equation}\label{Eq3.5}
D_{MKT}=\zeta U^{2} L.
\end{equation}

Both the hydrodynamic theory and the MKT have been successful in describing dynamic wetting phenomena (\cite{blake2006physics}). 
In realistic scenarios, it is plausible that dissipation is a combination of both mechanisms.
However, one can identify the dominant source of dissipation relevant to a specific problem. 
According to Equation (~\ref{Eq3.1}) and (~\ref{Eq3.5}):

\begin{equation}\label{Eq3.6}
\frac{D_{hyd}}{D_{MKT}}= \frac{\eta \cdot U^{2} \cdot \ln (\frac{R_{o}}{R_{i}})/ \theta_{D} \cdot L}{ \zeta \cdot U^{2} \cdot L }.
\end{equation}

For the PDMS$@$Au surface, the receding contact angle is $\theta_{rec} \approx $ \SI{96}{\degree}.
Assuming that the contact line velocity while receding is not too large, which is the case for highly viscous capillary bridges, $\theta_{D}\approx \pi/2$ is a reasonable estimate. 
Furthermore, on average, the ratio of viscosity and contact line friction is $\frac{\eta }{\zeta } \approx 80$ (\cite{duvivier2013toward}). 
Finally, assuming $\ln (R_{o}/R_{i})\approx 10$, a typical value from the literature (\cite{blake2006physics}), we obtain

\begin{equation}\label{Eq3.7}
\frac{D_{hyd}}{D_{MKT}} \approx 0.1.
\end{equation}

Therefore, it can be concluded that for such scenarios of wetting liquid bridges, i.e., rather slowly moving contact lines and hydrophobic surfaces, the dominant source of dissipation is most likely that of the MKT. 
However, in the following we attempt to treat the hydrodynamic theory and the MKT on an equal footing, using an approximate description. 
The hydrodynamic theory and the MKT suggest different relations between the dynamic contact angle and the contact line velocity. 
For example, in the simplified case of small contact angles, according to the hydrodynamic theory:

\begin{equation}\label{Eq3.8}
\theta _{D}^{3}= \theta _{m}^{3} + A \cdot Ca,
\end{equation}

and the MKT:

\begin{equation}\label{Eq3.9}
\theta _{D}^{2}= \theta _{S}^{2} + B \cdot Ca,
\end{equation}

where $\theta _{m}$ is the microscopic contact angle, $A$ and $B$ are some constants, and $Ca=\eta U/\gamma$  is the capillary number. 
Despite the differences in the scaling of the dynamic contact angle with $Ca$, a zeroth-order approximation for small capillary numbers yields the same value for the contact angle. 
Taking $\theta _{D} \approx$ $const.$, $D_{hyd}$ also scales as $U^2$. 
Therefore, we can combine the hydrodynamic wedge dissipation and the dissipation according to the MKT, introducing an effective contact line friction factor $\zeta _{CL}$

\begin{equation}\label{Eq3.10}
D_{CL}=\zeta _{CL} U^{2}L,
\end{equation}

where the subscript CL stands for contact line. Since, according to the estimate of Equation (~\ref{Eq3.7}), the energy dissipation according to the MKT is much larger than the hydrodynamic dissipation in the liquid wedge, the error due to considering a constant value for $\theta _{D}$ is minor. 
This means that even beyond the regime of small capillary numbers, the scaling of Equation (~\ref{Eq3.10}) is very likely a good approximation. 
Furthermore, it should be mentioned that Equation (~\ref{Eq3.2}) has been derived for hard substrates. 
Nevertheless, it has been shown that the dissipation on PDMS-pseudo brushes also scales as $U^{2}$ (\cite{rostami2023dynamic}). 
Therefore, the corresponding dissipation can also be included in the effective contact line friction. 
It is also necessary to identify other sources of dissipation and compare their magnitude to the energy dissipation at the three-phase contact line.
It can be shown that the dissipation according to the MKT is much larger than the viscous dissipation due to the radial flow in the bulk of the liquid. 
In that context, the term “radial flow” refers to the flow in the direction of contact-line motion. 
The corresponding shear stress on the solid surface can be approximated by $\eta U/h$ which acts on an area of approximately $d \cdot L$. 
Thus, the ratio of the viscous dissipation in the bulk due to the radial flow ($D_{rad}$) and the dissipation according to the MKT ($D_{MKT}$) can be written as follows:

\begin{equation}\label{Eq3.11}
\frac{D_{rad}}{D_{MKT}} \approx \frac{\eta \cdot U^{2}/h \cdot d \cdot L}{\zeta \cdot U^{2} \cdot L}.
\end{equation}

Substituting the aforementioned values, and considering the fact that for the surface used in this study at low contact line velocities $\theta _{D} \approx \pi/2$, or in other words, $h \sim d$, this ratio would be approximately $0.012$.
Therefore, it can be concluded that viscous dissipation in the bulk due to the radial flow is only of minor importance.

To complete the picture, the viscous dissipation in the bulk due to the axial flow ($D_{ax}$) needs to be estimated. Here, the term “axial flow” refers to the flow in the direction along which the liquid bridge extends. 
Similar to the above analysis for the radial direction, it can be shown that $D_{ax}\approx \eta \cdot U_{z}^{2}/h \cdot d\cdot L$, where $U_{Z}$ is some characteristic flow velocity in the axial direction. 
Taking the length of the contact line as an axial length scale and assuming $U$ to be a typical velocity in the radial direction, from the continuity equation, it can be shown that $U_{Z}\sim \frac{L}{d} U$.
Therefore,

\begin{equation}\label{Eq3.12}
D_{ax} \approx \left(\frac{L}{d} \right )^{2}\eta U^{2}L.
\end{equation}

The value $L/d$ is a geometric factor representing the slenderness of the liquid bridge. 
This ratio is a function of the surface wettability, liquid viscosity, and time. For example, a higher viscosity of the liquid and a larger contact angle on the surface result in an increased slenderness of the liquid bridge. 
In the context of this study, the maximum value of $L/d$ is around 3.5, which pertains to the breakup of the 70\% liquid on the PDMS$@$Au surface. 
This implies that $D_{ax}\sim 0.15D_{MKT}$, and thus $D_{ax}$ is non-negligible.

\begin{figure}
\centering
\includegraphics[width=0.75\linewidth]{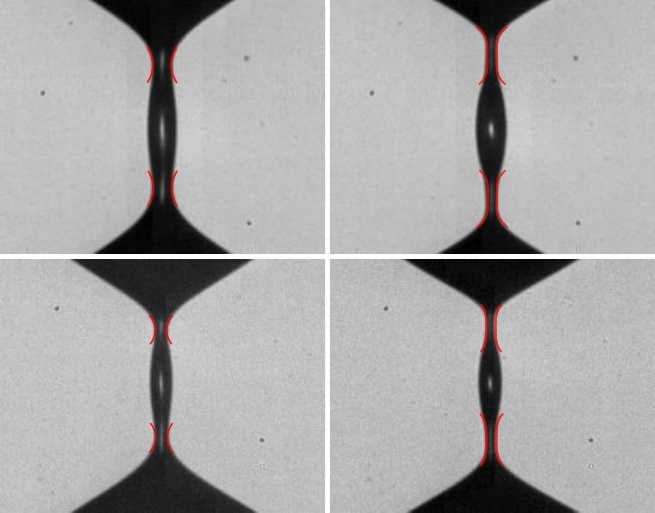}
\caption{Scaled profiles of the contact line in phase one (red curves) and corresponding instants in phase two (greyscale images). The top and bottom rows represent the 70\% and the 50\% liquid, respectively. The experiments were done on the PDMS$@$Au surface. The first column represents the instant in which the velocity is maximum in each phase, whereas the second column represents the last recorded image before the minimum width of a neck shifts from the middle to two new positions (again in each phase).}
\label{fig7}
\end{figure}

Despite the intricate dependency of $L/d$ on different parameters, if one considers the average over a broader time interval, this value remains approximately constant for a given liquid and substrate. 
This is due to undulation of the surface of the liquid bridge. 
When the capillary breakup begins, the minimum width of the bridge is in the middle. 
We call this period phase one. After some time has elapsed, the position of minimum width shifts to the two ends, creating two symmetrical necks and a bulge in the middle. 
We call this phase two. This phenomenon recurs and gives rise to multiple necks. 
The subsequent necks are geometrically similar to those in phase one. 
This can be made clear by comparing the profiles of the contact lines in the corresponding phases of the self-similar time evolution. 
Figure ~\ref{fig7} depicts the scaled profiles of the contact line in the constricted part of the liquid bridge in phase one (red curves) with the profiles at the equivalent instants in phase two (greyscale images) for two different liquids on the PDMS$@$Au surface.
The first column represents the instant where the velocity of the contact line is maximum in phase one and two, and the second column represents the last recorded image before the minimum width of a neck shifts from the middle to two new positions (again in each phase). 
As can be seen, the curves align reasonably well with the contact lines visible in the greyscale images. 
Thus, due to the geometric similarity between different phases, $L/d$ is constant when averaged over a longer time interval, leading to $D_{ax}\sim U^{2}$.
Therefore, the resulting dissipation due to all aforementioned channels can be summarized in the following equation by introducing a total friction factor $\zeta _{tot}$:

\begin{equation}\label{Eq3.13}
D_{tot} = \zeta _{tot} U^{2}L.
\end{equation}

Throughout this paper, we will refer to the combined effects of the mentioned mechanisms as the friction force. 
Nevertheless, it is helpful to know in which part of the parameter space which dissipation mechanism tends to dominate. 
In our experiments, we observed that by increasing the viscosity of the liquid or the contact angle of the surface, $L/d$ increases. 
The maximum value of $L/d$ (3.5) corresponds to the breakup of the 70\% liquid on the PDMS$@$Au surface, and the minimum value (1.9) corresponds to the 0\% liquid on the Au surface. 
It should be noted that many parameters such as surface roughness and contact angle hysteresis influence the wetting properties, and consequently the capillary breakup dynamics. 
However, to provide a simple characterization of the dominating dissipation mechanism, here we only consider the contact angle. 
Then, the following qualitative statement can be made:  For surfaces with smaller contact angle or liquids with lower viscosity, the major contribution to the total friction factor stems from contact line effects.

\subsubsection{Scaling arguments}\label{section3.2.2}

In the remainder of this section, a scaling relation is derived by considering the friction force and the capillary force. 
Considering the capillary bridge to be a slender thread, the curvature of the liquid surface is $\kappa \sim 1/h$ , see the close-up in Figure ~\ref{fig2}. 
The Laplace pressure due to this curvature, acting on an area $A$, balances the friction force:

\begin{equation}\label{Eq3.14}
\gamma \cdot \kappa \cdot A \sim \zeta _{tot} \cdot U \cdot L,
\end{equation}

where $U$ is a typical velocity in the radial direction. 
By taking $U\sim d/\tau $, and $A \sim h\cdot L$, the force balance reduces to

\begin{equation}\label{Eq3.15}
d\sim \frac{\gamma }{\zeta _{tot}}\tau .
\end{equation}

Equation (~\ref{Eq3.15}) is reminiscent of the well-known viscous law for the free capillary bridge, and this resemblance explains the similar qualitative behavior in the $d$ vs. $\tau$ diagrams. 
In the remainder of this paper, we will refer to the dynamic regime that follows the scaling of Equation (~\ref{Eq3.15}) as the viscous regime. 
In Figure ~\ref{fig6}.a, the dashed lines represent the asymptotes $d\sim \tau$, obtained as follows: first, a power law with a power of $1$ is fitted to the 70\% liquid in the region \SI{6}{\micro \meter} $< d < $ \SI{40}{\micro \meter}. 
Then, the rest of the fit functions (dashed lines) are calculated by considering the viscosity dependence of the friction factor $\zeta _{tot}\sim \eta $. 
It should be noted that the intricate dependence of $\zeta _{tot}$  on surface tension was neglected when computing these fit functions, i.e. only the viscosity dependence was considered. 
Instead, an average value of the surface tension (\SI{69.23}{\milli \newton \per \meter}) was used.
The lower bound of $d$ in curve fitting is chosen according to the optical resolution of the imaging system, while the upper bound corresponds to the time where the liquid bridge of roughly uniform width appears. 
In the context of the capillary breakup of free liquid bridges, it has been established that the entrance to the viscous scaling regime is a function of the geometry of the system or the shape of the liquid bridge and independent of the viscosity of the liquid (\cite{li2016capillary}). 
We have observed a similar behavior for the wetting liquid bridges. 
This is verified by the fact that for the liquids that follow the scaling fairly well, namely the 50-70\% liquids, the power-law behavior starts at $d\approx$ \SI{40}{\micro \meter}. 
Nevertheless, deviations occur for the 30\% liquid, which can be attributed to the increasing importance of inertia at higher velocities. 
Figure ~\ref{fig6}.b shows the scaled minimum width ($\overline{d }$) vs. scaled time before breakup ($\overline{\tau }$). 
Here, the bridge width is scaled by dividing by some length scale $a$, and time is scaled with $\gamma a/\zeta _{tot}$. 
Furthermore, a physically meaningful length scale will be introduced in Section ~\ref{section3.3}. 
For the moment, the length scale is arbitrarily chosen to be \SI{75}{\micro \meter} (the minimum width where the 70\% liquid has the maximum velocity). 
The collapse of the curves for the 50-70\% liquids indicates that they are indeed in the viscous regime. 
A natural velocity scale for the problem is $U^\ast=\gamma /\zeta_{tot}$, which is similar to the intrinsic capillary velocity ($\gamma /\eta $). 
Figure ~\ref{fig6}.c shows the scaled breakup velocity ($U/U^\ast$) as a function of the scaled minimum width of the capillary bridge. 

\subsection{Influence of geometric constraints}\label{section3.3}

\begin{figure}%
    \centering
    \subfloat[\centering ]{{\includegraphics[width=6cm]{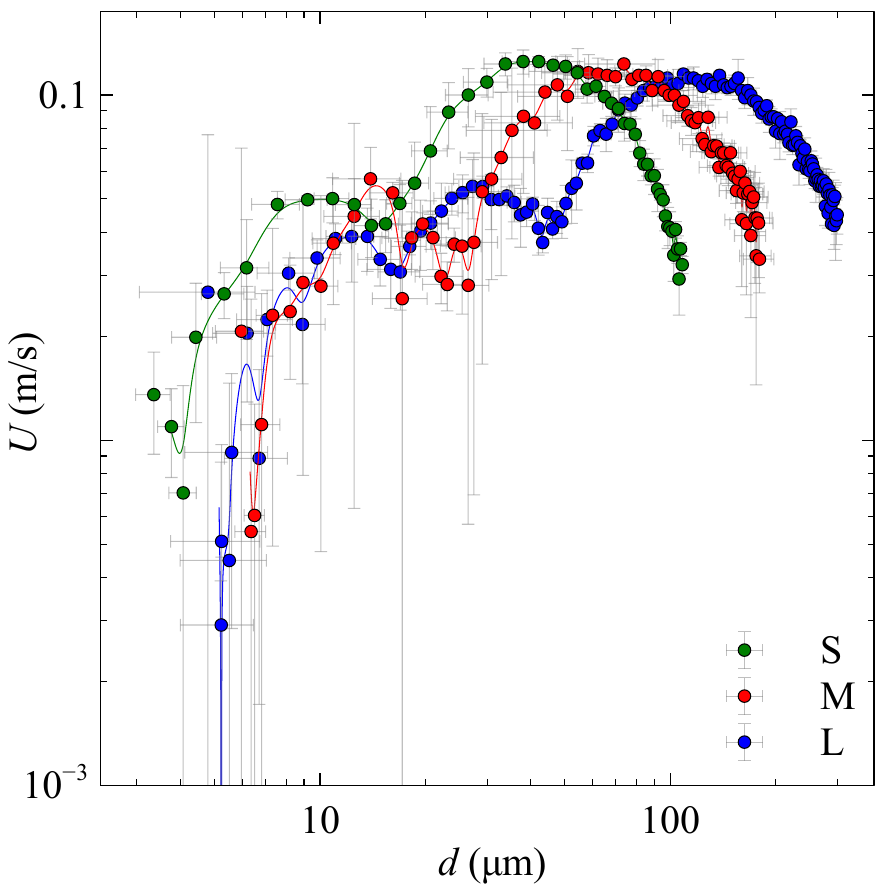} }}%
    \qquad
    \subfloat[\centering ]{{\includegraphics[width=6cm]{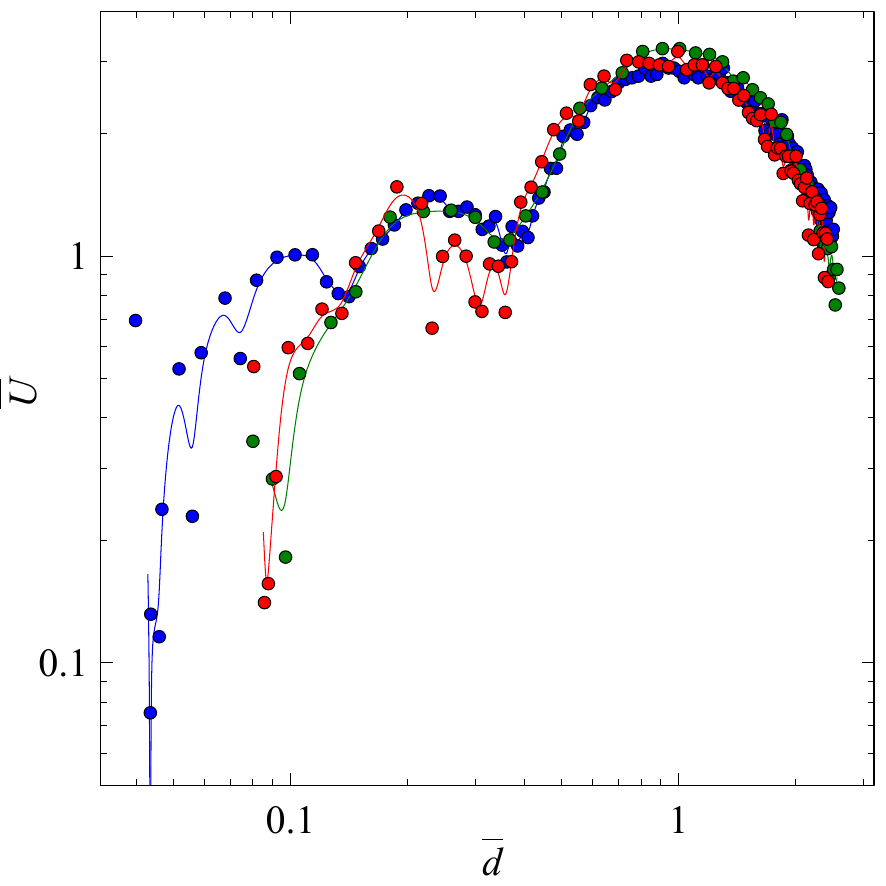} }}%
    \qquad
    \subfloat[\centering ]{{\includegraphics[width=6cm]{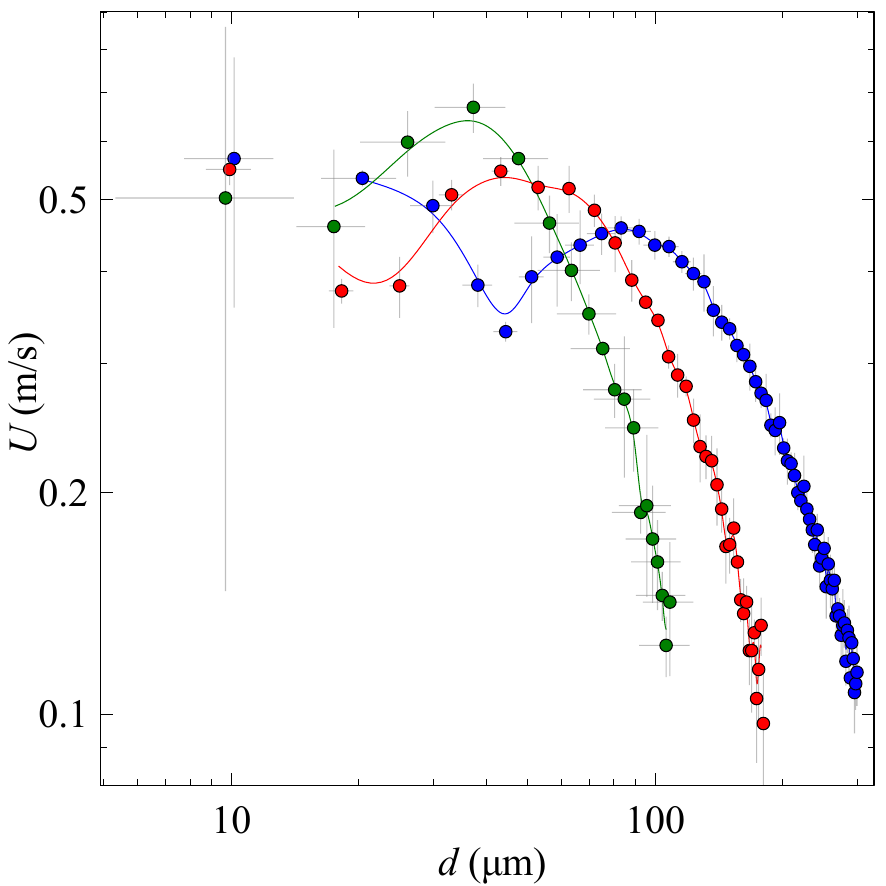} }}
        \qquad
    \subfloat[\centering ]{{\includegraphics[width=6cm]{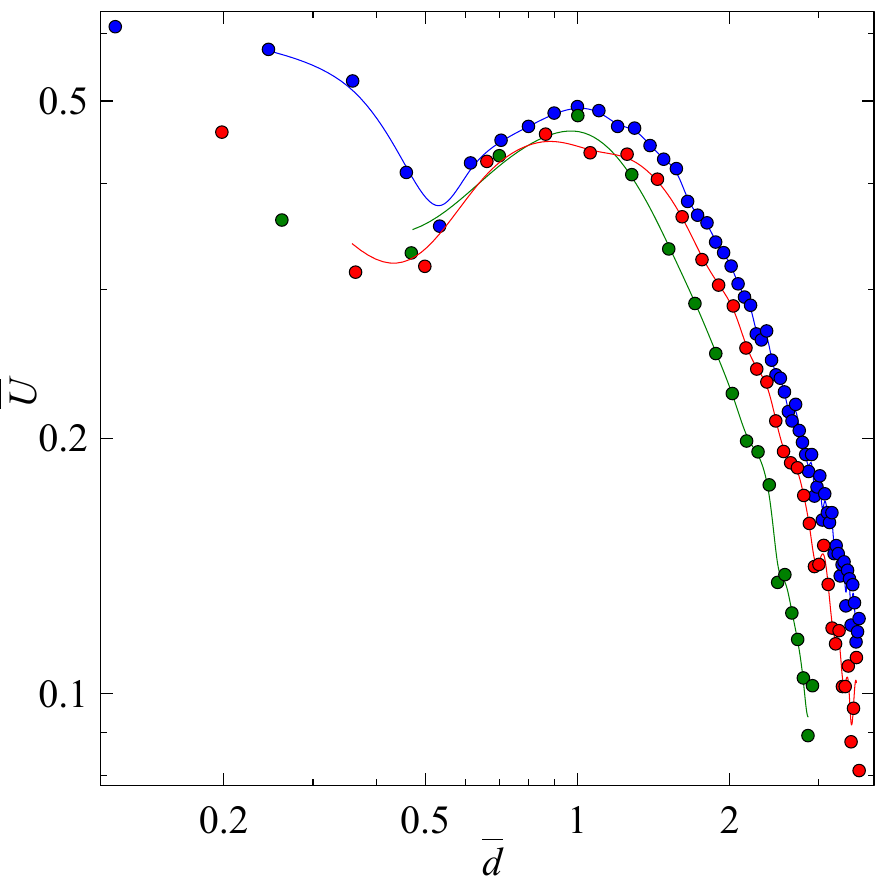} }}%
    \caption{Influence of geometric constraints on the dynamic behavior of the capillary bridge: breakup velocity vs. minimum width for (a) the 60\% liquid and (c) the 0\% liquid. The error bars represent the standard deviation. Scaled breakup velocity vs. scaled minimum width for (b) the 60\% and (d) the 0\% liquid. The legend of subfigure (a) applies to all subfigures.}%
    \label{fig8}%
\end{figure}

By altering the gap width ($W$ in Figure ~\ref{fig1}) of the PDMS block, the effect of external length scales on the breakup dynamics can be studied. 
To achieve this, experiments have been conducted with three different gap widths: \SI{3}{\milli \meter}, \SI{1.5}{\milli \meter}, and \SI{0.75}{\milli \meter}, denoted as $\textup{L}$, $\textup{M}$, and $\textup{S}$, respectively. 
$U$ vs. $d$ plots for the 60\% and 0\% liquids can be found in Figure ~\ref{fig8}.a and  Figure ~\ref{fig8}.c, respectively. 
The key difference between the plots is the fact that by decreasing the size of the system, there is a significant increase in the maximum breakup velocity for the 0\% liquid, whereas for the 60\% liquid, there is only a shift along the x-axis and no apparent change in the maximum velocity, see  Figure ~\ref{fig8}.a and Figure ~\ref{fig8}.c. 
This difference can be explained by considering the forces at play in each regime. For the 0\% liquid (inertial regime), capillary forces are balanced by inertial forces, while for the 60\% liquid (viscous regime), they are balanced by friction forces.
Typical time and velocity scales for each of these regimes can be written as follows:

\begin{equation}\label{Eq3.16}
\hat{t}_{v}=\frac{\zeta _{tot}a}{\gamma }, \hat{d}=a \Rightarrow \hat{V}_{v}=\frac{\gamma }{\zeta_{tot}},
\end{equation}

\begin{equation}\label{Eq3.17}
\hat{t}_{i}=\left(\frac{\rho a^{3}}{\gamma } \right )^{0.5}, \hat{d}=a \Rightarrow \hat{V}_{i}=\left(\frac{\gamma}{\rho a } \right )^{0.5},
\end{equation}

where $\rho$ is the mass density, $a$ is a characteristic length scale, and the subscripts $i$ and $v$ indicate the inertial and viscous regimes, respectively. 
Evidently, in the viscous regime, the characteristic velocity remains independent of any length scale, while in the inertial regime, the velocity scales inversely with the square root of the length scale.
The fact that by increasing the size, the dynamics slows down should not be surprising for inertial dynamics.

To proceed, we need to identify the characteristic length scale $a$, which can depend on the geometric scales of the problem, but also on combinations of other problem parameters with the dimension of a length. 
For the latter, the capillary length could come into play, but can be ruled out because of the negligible relevance of gravitational forces on sub-millimeter scales. 
With respect to the geometric scales, in addition to the gap width, other scales such as the channel width could potentially influence the dynamics of the capillary bridge. 
Therefore, the relationship between $W$ and $a$ is not immediately clear. 
However, in any case, there is a minimum width of the capillary bridge where the bridge becomes unstable. 
This minimum width is called $a_{crit}$. 
Capillary bridges formed under different geometric constraints but in the same dynamic regime are expected to follow a universal dynamic behavior. 
In other words, with the correct scaling, the corresponding curves in the $U$ vs. $d$ space should collapse onto a master curve. 
Since by definition $a_{crit}$ indicates the starting point of the instability, one can hypothesize that choosing $a=a_{crit}$ and the velocity scales of equations (~\ref{Eq3.16}) and (~\ref{Eq3.17}) would result in a collapse of curves. 
However, determining $a_{crit}$ proves to be challenging. 
To eliminate these complications, the width of the capillary bridge at the maximum velocity ($a_{max}$) is taken as the length scale. 
Figure ~\ref{fig8}.b shows an excellent collapse of curves for the 60\% liquid based on this length scale. 
It can be seen in the $U$ vs. $d$ curves that the onset of instability for highly viscous liquids occurs approximately at $3a_{max}$, which indicates that, to a reasonable approximation, there is a fixed relationship between $a_{crit}$ and $a_{max}$.
Correspondingly, the scalings with $a_{crit}$ and $a_{max}$ appear to be equivalent.

\subsection{Formation of satellite droplets} \label{section3.4}

The purpose of this section is to qualitatively compare the morphologies of the final configurations after liquid-bridge breakup as observed for free and wetting liquid bridges. 
As can be seen in Figure ~\ref{fig3} and Figure ~\ref{fig5} and as discussed before, the position of the minimum width of a capillary bridge does not always remain in the middle; instead, it shifts towards the two ends.
This phenomenon occurs repeatedly, leading to the formation of several necks in the liquid bridge, and consequently, the generation of multiple satellite droplets (\cite{bhat2010formation, tomotika1935instability}). 
For a free capillary bridge surrounded by an inviscid fluid, the following behavior is observed in the high- and low viscosity limit. 
In viscous dynamics, capillary bridges tend to form an elongated thread where the minimum radius remains in the middle for a long time, whereas bridges of low-viscosity soon tend to develop necks close to the two ends, resulting in a comparatively large satellite droplet in the middle. 
For highly viscous liquids, viscous dynamics describes the dynamic behavior throughout most of capillary breakup. 
However, the final stage of the breakup occurs in the viscous-inertial regime (\cite{li2016capillary}). 
In the viscous-inertial regime of capillary breakup, there is a small satellite droplet in the middle of the bridge, with two thin threads on either side.

\begin{figure}
\centering
\includegraphics[width=0.75\linewidth]{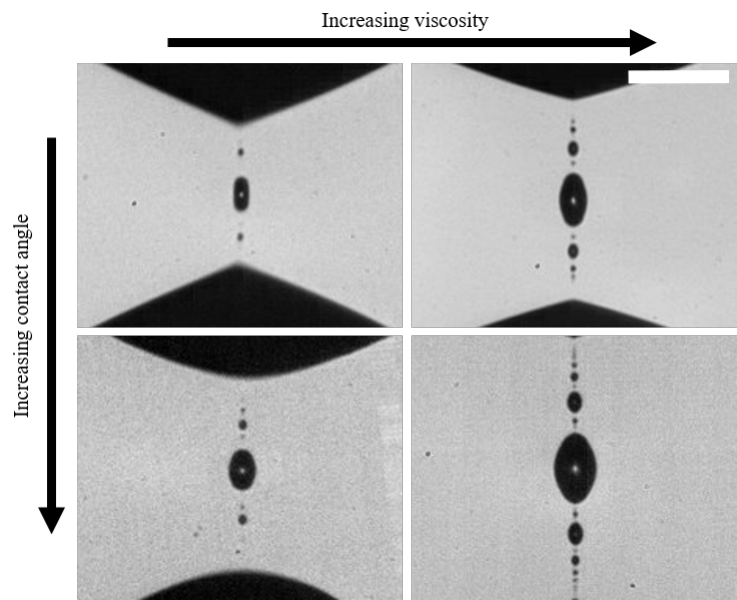}
\caption{Change in number and shape of the satellite drops with viscosity and the contact angle of the substrate for wetting capillary bridges. The scale bar represents \SI{100}{\micro \meter}.}
\label{fig9}
\end{figure}

For a wetting liquid bridge, the number of necks before breakup is a function of the surface properties as well as liquid viscosity. 
Analogous to a free liquid bridge surrounded by an inviscid fluid, the highly deformed shape is also observed for lower viscosities (see, e.g., the last frame showing the 0\% liquid in Figure ~\ref{fig5}). 
However, unlike a free liquid bridge, the satellite droplet in the middle is larger for higher viscosities. 
Additionally, within the chosen parameter range in this study, longer capillary bridges ‒whether due to higher viscosity of the liquid or larger contact angle of the surface‒ produce more satellite droplets, as can be seen in Figure ~\ref{fig9}. 
As mentioned earlier, many parameters of the surface influence the wetting behavior and thus the breakup dynamics or the morphology of the satellite droplet.
However, for the purpose of conciseness, here we only refer to the contact angle.

\section{Conclusion}

The breakup dynamics of viscous capillary bridges on solid surfaces was studied experimentally. 
It was shown that the surface properties play a significant role in the dynamics, as indicated by the fact that the breakup process is much faster for the PDMS$@$Au surface compared to Au surface. 
The solid surface plays a twofold role. 
First, it slows down the dynamics of the capillary bridge through friction forces at the three-phase contact line. 
Contact-line friction is the dominant dissipation mechanism at small contact angles and low viscosities. 
Second, also viscous dissipation at the surface away from the contact line plays a role. This contribution tends to become important with increasing liquid viscosity and contact angle.

In the experiments, the viscosity of the working liquids (water/glycerol mixtures) was systematically varied. 
As expected, the breakup velocity decreases with increasing viscosity. 
It was shown that the effects of molecular origin according to the MKT likely dominate over hydrodynamic dissipation in the liquid wedge. 
Furthermore, when the liquid bridge becomes very long, also the viscous dissipation due to the axial flow becomes substantial. 
To interpret the data, we introduced a total friction factor that incorporates the effects of molecular origin according to the MKT, hydrodynamics in the liquid wedge, and the viscous forces due to axial flow.

The most important quantity observed in the experiments on capillary bridge breakup is the minimum width of the bridge as a function of time.
In the framework of the total friction force, a scaling model for the time dependence of the minimum width was formulated. 
The viability of this scaling is demonstrated by the fact that in the viscous regime, the curves showing the minimum width vs. time collapse on the same master curve upon rescaling of the axes.
Furthermore, the influence of external geometrical constraints on the breakup dynamics was studied. 
It was shown that changing the external length scale does not affect the maximum breakup velocity of the viscous capillary bridges. 
By contrast, for inertia-dominated capillary bridges, the maximum velocity increases as the size of the system decreases. 
It was hypothesized that the length scale governing the capillary breakup is the minimum width of the capillary bridge at the onset of instability.
However, due to the difficulties in determining this value precisely, it was shown that the minimum width at the maximum velocity could serve as an equally viable length scale for the problem. 
This was confirmed by rescaling the axes in a diagram that shows the breakup velocity vs. time for different external length scales.

Lastly, a qualitative assessment of the morphologies of the satellite droplets after capillary breakup revealed similarities and differences between free and wetting bridges. 
Similar to free liquid bridges, low viscosity wetting bridges experience a highly deformed shape before breakup. 
However, unlike free liquid bridges, wetting bridges with higher viscosity produce a larger satellite droplet in the middle. 
Increasing the contact angle on the surface and the liquid viscosity leads to longer bridges and more satellite droplets.

\section{Acknowledgements}

We express our gratitude to Alexander Erb for his invaluable assistance with the experimental part.

\section {Funding}

This study was funded by German Research Foundation (DFG) within the Collaborative Research Centre 1194 “Interaction of Transport and Wetting Processes,” Project- ID 265191195, subprojects A02a and A02b

\section {Declaration of interests}
The authors report no conflict of interest.

\bibliographystyle{jfm}
\bibliography{jfm}


\end{document}